\documentclass[conference]{IEEEtran}
\IEEEoverridecommandlockouts
\usepackage{cite}
\usepackage{amsmath,amssymb,amsfonts}
\usepackage{algorithm} 
\usepackage{algpseudocode} 
\usepackage{graphicx}
\usepackage{textcomp}
\usepackage{xcolor}
\usepackage{float,color,comment}
\usepackage{amsmath}
\usepackage{caption}
\usepackage{subfigure}
\usepackage{graphics} 
\usepackage{epsfig} 
\usepackage{multirow}
\graphicspath{{figure/}}
\def\BibTeX{{\rm B\kern-.05em{\sc i\kern-.025em b}\kern-.08em
    T\kern-.1667em\lower.7ex\hbox{E}\kern-.125emX}}

\usepackage{array}

\begin{document}

\title{Learning-Based MIMO Channel Estimation under 
Spectrum Efficient Pilot Allocation and Feedback \\
\thanks{M. del Rosario and Z. Ding are with the Department of Electrical and Computer Engineering, University of California at Davis, Davis, CA 95616 USA (e-mail: mdelrosa@ucdavis.edu, zding@ucdavis.edu).}
}

\author{\IEEEauthorblockN{Mason del Rosario and Zhi Ding} }

\maketitle

\begin{abstract} 
Wireless links using
massive MIMO transceivers are vital for next generation
wireless communications networks networks.
Precoding in Massive MIMO transmission 
requires accurate downlink
channel state information (CSI). 
Many recent works have effectively applied deep learning (DL)
to jointly train UE-side compression networks for delay domain CSI 
and a BS-side decoding scheme.
Vitally, these works assume that the full delay domain
CSI is available at the UE, but 
in reality, the UE must estimate the delay domain
based on a limited number of frequency domain pilots.
In this work, we propose a linear pilot-to-delay (P2D) estimator 
that transforms sparse frequency pilots 
to the truncated delay CSI. We show that 
the P2D estimator is accurate under
frequency downsampling, and 
we demonstrate that the P2D estimate can
be effectively utilized with existing autoencoder-based CSI estimation networks.
In addition to accounting for pilot-based estimates of downlink CSI,
we apply unrolled optimization networks to emulate iterative
solutions to compressed sensing (CS), and we demonstrate better estimation 
performance than prior autoencoder-based DL networks. 
Finally, we investigate the efficacy of trainable CS networks for in a
differential encoding network for time-varying CSI estimation, 
and we propose a new network, MarkovNet-ISTA-ENet, comprised of both a CS network for initial CSI estimation and multiple autoencoders to estimate the error terms. We demonstrate that this heterogeneous network has better
asymptotic performance than networks comprised of only one type of network.

\end{abstract}


\section{Introduction}

The modern wireless networks utilizing massive
multiple-input multiple-output (MIMO) technologies are critical to 
achieving high link capacity \cite{ref:mimo-capacity}. 
To realize these capacity gains, 
MIMO base stations (gNB) require accurate downlink
channel state information (CSI) 
for user equipment (UE) precoding.
While uplink-downlink reciprocity in TDD systems \cite{ref:Kaltenberger2010relative, ref:mi2017massive, ref:Gao2010utilization}
can be exploited to estimate downlink CSI via uplink
CSI, FDD networks exhibit comparatively weak channel reciprocity.
Thus, feedback from UEs is necessary for downlink CSI estimation
and UE-specific precoding at the BS.

Many recent works have applied deep learning (DL)-based CSI compression and 
estimation including the successful application of 
convolutional neural networks (CNNs) in autoencoder \cite{ref:csinet, ref:Lu2020CRNet, ref:Hussien2020PRVNet, ref:Sun2020AnciNet}, 
the integration of uplink magnitude-reciprocity 
at the decoding CNN \cite{ref:dualnet},
and the exploitation of temporal CSI coherence \cite{ref:csinet-lstm, ref:Liu2020MarkovNet}. 

Broadly speaking, many works assume that the full downlink CSI matrix is available at the UE. In practice, this assumption is not met since individual elements of the downlink CSI are estimated using pilot symbols. Pilot estimation effectively means that the CSI available at the UE is a sparse, downsampled version of the full CSI. Assuming pilot estimation, DL-based downlink CSI compression and estimation schemes must account for input data that are low-resolution and noise corrupted compared to the full, ground truth CSI data.

To effectively utilize pilot-based CSI at the UE for downlink CSI estimation, 
this work presents the following contributions:

\begin{itemize}
	\item \textbf{Pilots-to-Delay (P2D)}: Based on a limited number of pilot-based estimates, we propose an accurate estimator of the truncated delay-domain CSI at the UE. Using this estimator as the input to a range of deep learning-based CSI compression networks, we show that this estimator provides a suitable surrogate for ground-truth delay domain CSI. To conform to 3GPP specifications, we outline the pilot allocation in the time-frequency resource grid based on CSI-RS. 
	\item \textbf{Pilot-based Differential Encoding}: Using the proposed P2D estimates at the UE, we propose to encode and feed back the estimation error. To compress the error terms, we compare unrolled optimization networks, which enable trainable compressive sensing algorithms via deep learning, with autoencoder networks, which have been commonly used in CSI feedback literature. We show that a differential network combining both unrolled compressed sensing networks and autoencoders can outperform prior autoencoder-based approaches to differential encoding.
\end{itemize}

\section{System Model} \label{sec:system-mod}
Without loss of generality, we consider a single-cell massive MIMO system
with $N_b \gg 1$ antennas at the gNB serving multiple UEs, each 
with a single antenna. 
The network operates under orthogonal frequency division multiplexing (OFDM) 
with $N_f$ subcarriers. On the downlink, the received UE signal on the $m$-th subcarrier/subband is
\begin{equation}
	y_{m} =\mathbf{h}_{m}^H\mathbf{w}_{m}x_{m} + n_{m}, 
\label{equ1}
\end{equation}
where $\mathbf{h}_{m} \in \mathbb{C}^{N_b\times1}$ is the
downlink CSI 
of the $m$-th subcarrier,
$\mathbf{w}_{m} \in \mathbb{C}^{N_b\times1}$ denotes the precoding vector, $x_{m}\in \mathbb{C}$ is the transmitted symbol, 
and $n_{m}\in \mathbb{C}$ is additive noise, and
$(\cdot)^H$ denotes conjugate transpose.
The downlink CSI matrix in the spatial-frequency domain 
is $\mathbf{H} \in \mathbb{C}^{N_b\times N_f}$. 


\begin{figure*}[!hbtp]
    \centering
    {
      \fontsize{8pt}{8pt}
      \def\svgwidth{1.0\linewidth}
      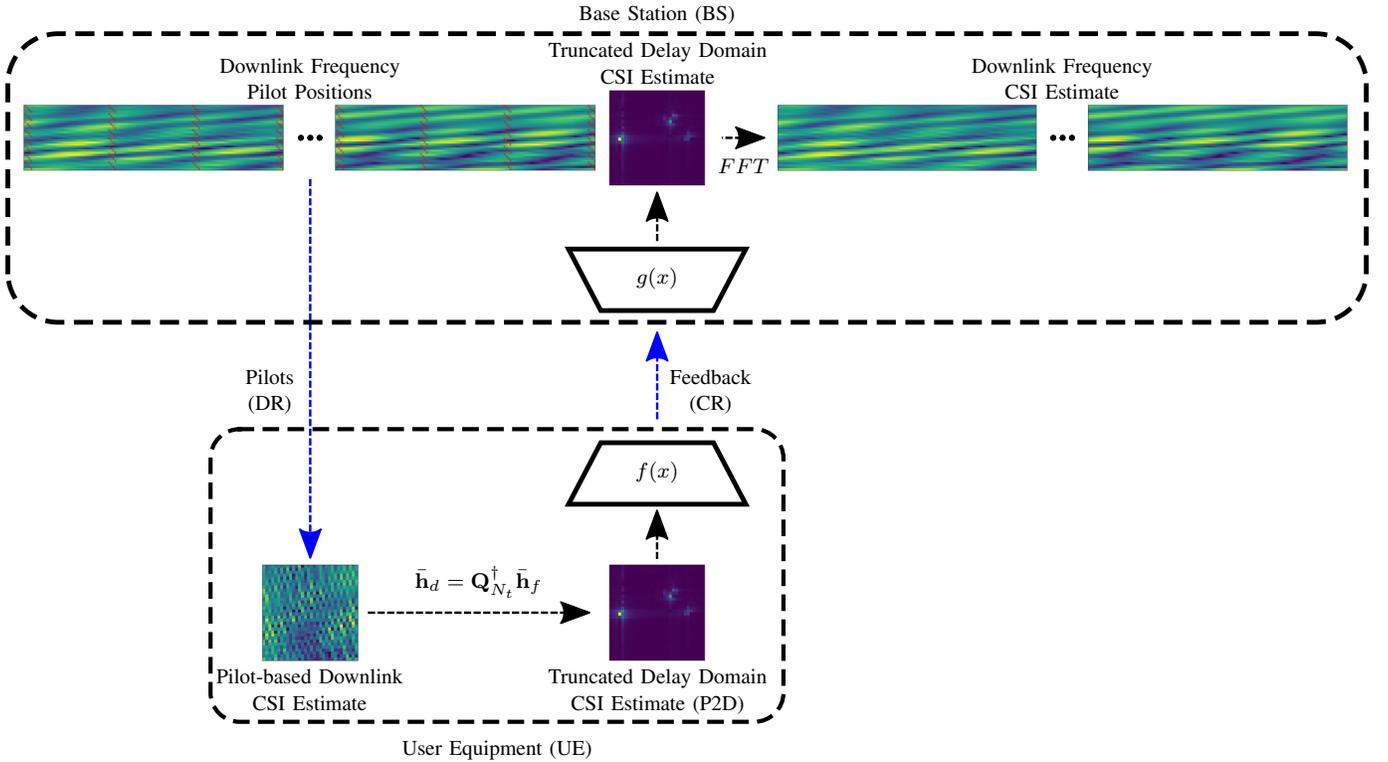
    }
    \caption{Compressive CSI estimation based on the linear P2D estimator. First, downlink pilots are used to estimate a sparse, downsampled frequency domain CSI estimate of size $M_f << N_f$. Then, the P2D estimator, $\mathbf{Q}^\dag_{N_t}$ of (\ref{eq:p2d_short}), is applied to establish an accurate truncated delay domain CSI estimate. After P2D estimation, a learnable UE (BS) transform $f(x)$ ($g(x)$) is used to compress and decode the feedback, respectively. Finally, the delay domain estimate can be transformed into the frequency domain for precoding.}
    \label{fig:p2d}
\end{figure*}

To estimate the downlink CSI, pilots are used at small number of spatial-frequency locations, resulting in a downsampled version of the CSI matrix $\bar{\mathbf{H}} \in \mathbb{C}^{M_b\times M_f}$ where $M_f < N_f$ and $M_b < N_b$. To construct $\bar{\mathbf{H}}$, denote each antenna vector, $\tilde{\mathbf{h}}_{a}\in\mathbb{C}^{N_f}$, which is obtained using a pilot frequency selection matrix, $\mathbf{P} \in [0,1]^{M_f \times N_f}$ where each row is a one-hot vector. Thus, the matrix-vector product $\bar{\mathbf{h}}_{j} = \mathbf{P}\tilde{\mathbf{h}}_j\in\mathbb{C}^{M_f}$ represents a downsampled version of the antenna vector $\tilde{\mathbf{h}}_{a}$. 

While the discussion so far has focused on spatial-frequency CSI, most works in compressive CSI estimation opt to use the angular-delay domain CSI, $\tilde{\mathbf{H}}$, which exhibits greater sparsity and is more amenable to compression than the spatial-frequency domain \cite{ref:csinet}. Such works presume that the angular-delay domain data are readily available at the UE, but in reality, the UE only has access to the CSI estimated via pilots (i.e., the downsampled frequency-spatial CSI matrix $\bar{\mathbf{H}}$).




\section{Linear Prediction of Delay-domain CSI via Frequency-domain Pilots} \label{sec:pilots-to-delay}

\begin{figure*}[!hbtp]
    \centering
    {
      \fontsize{8pt}{8pt}
      \def\svgwidth{1.0\linewidth}
      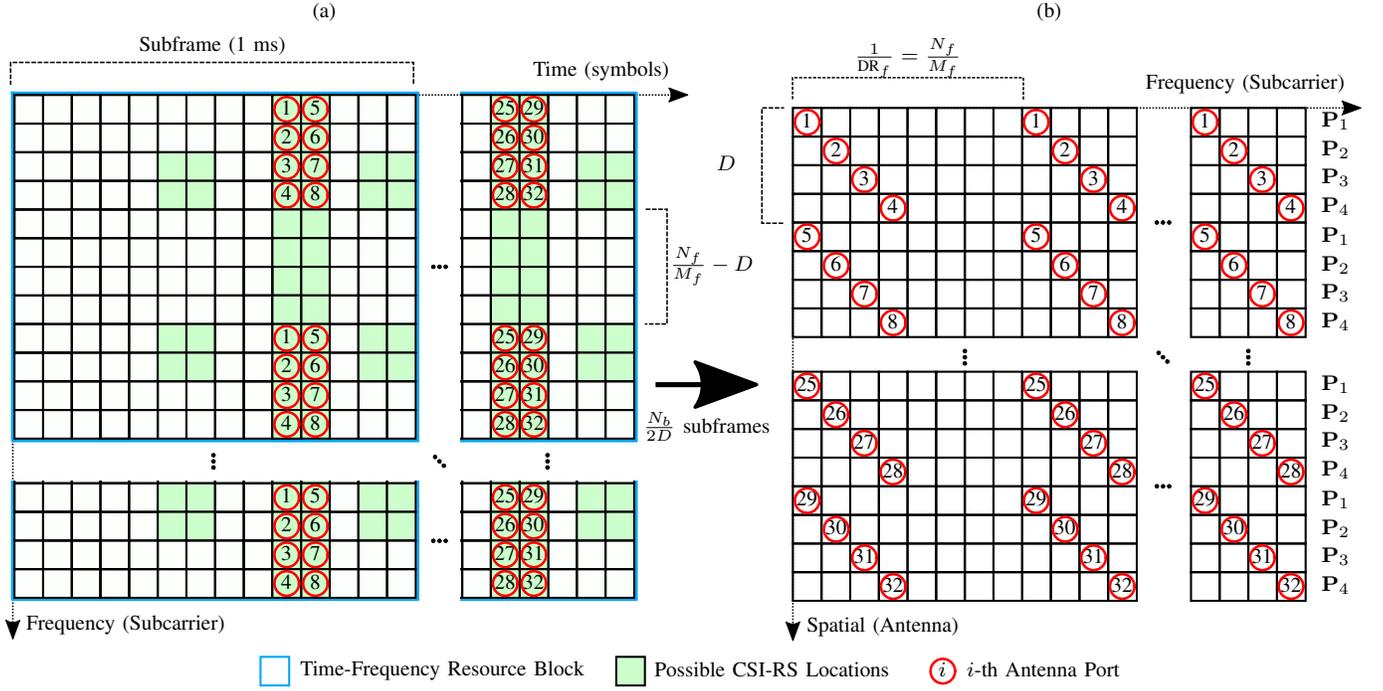
    }
    \caption{(a) LTE Resource Blocks and CSI-RS locations where antenna port pilots are allocated. (b) Schematic for diagonal pilots with relevant parameters, size of diagonal $D$ and frequency downsampling ratio $\text{DR}_f$. In this diagram, $N_b=32, D=4, \text{DR}_f=\frac 18$. The pilot matrix $\mathbf{P}_j$ indicates the downsampling pattern for the $j$-th element of the diagonal pattern. The number of subframes necessary to populate (b) is inversely proportional to $D$.}
    \label{fig:p2d_diag}
\end{figure*}


Using the limited number of frequency domain pilots available at the UE, we can estimate the truncated delay domain data. This delay domain estimate is directly compatible with the commonly used CSI basis in prior deep learning based CSI compression works \cite{ref:csinet, ref:Liu2020MarkovNet}, which have demonstrated high estimation accuracy under substantial compression.

\subsection{Frequency Domain Downsampling}

Donsider the case where downsampling is performed along the frequency axis such that $M_f$ subcarriers of the original $N_f$ subcarriers remain. Downsampling is done by applying the pilot matrix $\mathbf{P} \in \mathbb{C}^{M_f\times N_f}$ to the frequency domain vector $\mathbf{h}_f$, resulting in the pilot vector $\bar{\mathbf{h}}_f \in \mathbb{C}^{M_f}$. Note that $\mathbf{h}_f$ is one of the rows of the spatial-frequency matrix $\mathbf{H}$.

To relate the frequency and delay domain, denote the Fast Fourier Transform (FFT) dual,
\begin{align*}
	\mathbf{F}_{N_f}\mathbf{h}_d &= \mathbf{h}_f.
\end{align*}
Applying the pilot matrix to both sides, we have
\begin{align*}
	\mathbf{P}\mathbf{F}_{N_f}\mathbf{h}_d &= \mathbf{P}\mathbf{h}_f \\
	\mathbf{Q}\mathbf{h}_d &= \bar{\mathbf{h}}_f,
\end{align*}
where $\mathbf{Q}=\mathbf{P}\mathbf{F}_{N_f}$. Given the sparsity of CSI in the delay domain, we may truncate $\mathbf{Q}$ to the first $N_t$ columns and restrict our attention to the truncated delay domain vector, $\bar{\mathbf{h}}_{d} \in \mathbb{C}^{N_t}$,
\begin{align*}
	\mathbf{Q}_{N_t}\bar{\mathbf{h}}_d &= \bar{\mathbf{h}}_f.
\end{align*}
To solve for $\bar{\mathbf{h}}_d$, we perform the pseudoinverse $\mathbf{Q}_{N_t}^\dag$,
\begin{align}
	\bar{\mathbf{h}}_d &= \mathbf{Q}^\dag_{N_t}\bar{\mathbf{h}}_f \label{eq:p2d_short} \\
	&= (\mathbf{Q}_{N_t}^{H}\mathbf{Q}_{N_t})^{-1}\mathbf{Q}_{N_t}^{H}\bar{\mathbf{h}}_f \label{eq:p2d_long}
\end{align}
This solution relies solely on the downsampling matrix, $\mathbf{P}$, and the FFT matrix, $\mathbf{F}_{N_f}$. We call this solution the pilots-to-delay (P2D) estimator since it allows us to estimate the truncated delay domain CSI ($\mathbf H$) based on sparse frequency domain pilots. Figure~\ref{fig:p2d} shows where the P2D estimator fits into the overall CSI feedback and estimation process.

In contrast with a ``Compression Ratio (CR)" that is typically reported in the feedback stage, the P2D estimator is associated with a ``Frequency Downsampling Ratio ($\text{DR}_f$)," which is given as
\begin{align}
    \text{Frequency Downsampling Ratio ($\text{DR}_{f}$)} &= \frac{M_f}{N_f}.
\end{align}

\subsection{Diagonal Pilot Patterns for LTE Compatibility}

In the LTE specification, downlink pilots for antenna ports are allocated to specific resource elements (CSI-RS) in the time-frequency resource grid \cite{ref:Asplund2020}. For a MIMO array, the different antenna ports are allocated to CSI-RS locations in the resource grid, and multiple subframes might be necessary to acquire the entire downsampled CSI matrix. The number of subframes necessary depends on two design parameters: 1) the size of the diagonal pattern, $D$, and 2) the frequency downsampling ratio, $\text{DR}_f$.

\begin{algorithm}
	\caption{Pilots to Delay (P2D) Estimator for Diagonal Pilot Pattern} 
	\label{alg:p2d-diag}
	\begin{algorithmic}[1]
	    \State \textbf{\emph{Input}}: Pilot spatial-frequency CSI, $\bar{\mathbf{H}}\in\mathbb{C}^{N_b\times M_f}$
	    \State \textbf{\emph{Input}}: Diagonal parameter, $D$
	    \State \textbf{\emph{Initialize}}: Spatial-delay CSI, $\bar{\check{\mathbf{H}}}\in\mathbb{C}^{N_b\times N_t}$
	    \State \textbf{\emph{Initialize}}: Angular-delay CSI, $\bar{\tilde{\mathbf{H}}}\in\mathbb{C}^{N_b\times N_t}$
		\For {$i=1,2,\ldots, N_b$}
		    \State \textbf{\# \emph{Index for $j$-th pilot matrix}}
		    \State $j = ((i-1) \text{ mod } D) + 1$
		    \State \textbf{\# \emph{Apply P2D to $i$-th pilot subcarrier}}
		    \State $\bar{\mathbf{h}}_f(i)=\bar{\mathbf{H}}(i,:)$ 
			\State $\bar{\check{\mathbf{H}}}(i,:) = \mathbf{Q}^\dag_{j,N_t}\bar{\mathbf{h}}_f(i)$
		\EndFor
	    \State \textbf{\# \emph{Convert spatial domain to angular}}
		\State $\bar{\tilde{\mathbf{H}}} = \text{FFT}(\bar{\check{\mathbf{H}}})$ 
		\State \textbf{\emph{Return}} $\bar{\tilde{\mathbf{H}}}$
	\end{algorithmic} 
\end{algorithm}

\begin{figure*}[!hbtp]
    \centering
    {
      \fontsize{8pt}{8pt}
      \def\svgwidth{1.0\linewidth}
      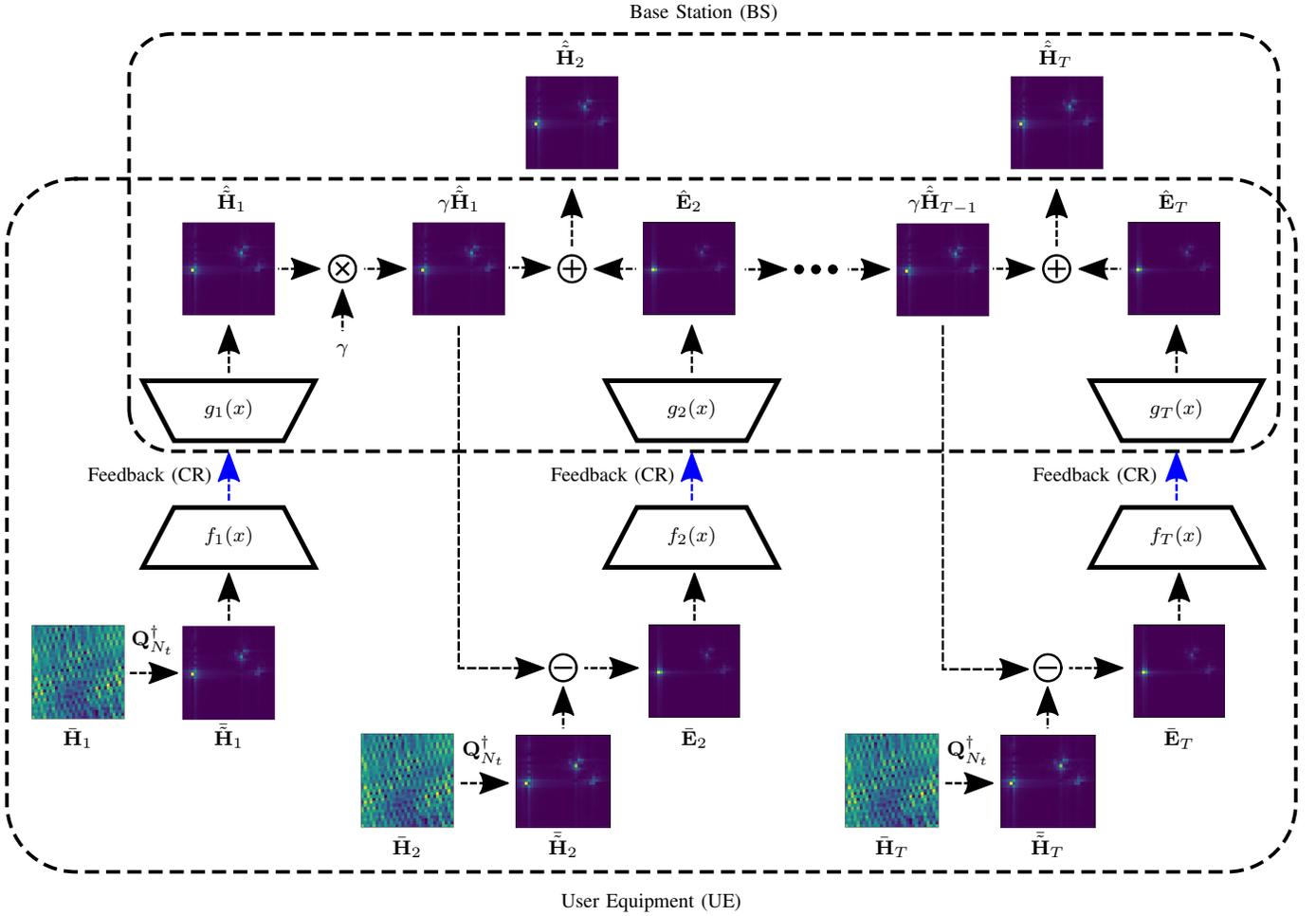
    }
    \caption{Diagram of a CSI estimation network using compressed differential feedback based on the linear P2D estimator. First, downlink pilots are used to estimate a downsampled frequency domain CSI estimate, $\bar{\mathbf{H}}_t\in\mathbb{C}^{N_b \times M_f}$ where $M_f << N_f$ at the $t$-th timeslot. Then, the P2D estimator, $\mathbf{Q}^\dag_{N_t}$ of (\ref{eq:p2d_short}), is applied to estimate $\tilde{\mathbf{H}}_t$. After P2D estimation, the learnable transforms $f_t(x)$ and $g_t(x)$ are used to compress and decode the feedback, respectively. For $t=1$, the encoder/decoder are applied directly to $\tilde{\mathbf{H}}_1$. In all subsequent timeslots ($t > 1$), the differential term $\mathbf{E}_t$ is compressed and fed back.}
    \label{fig:markov-p2d}
\end{figure*}

Figure~\ref{fig:p2d_diag}a illustrates our proposed pilot allocation for an LTE time-frequency resource grid, and Figure~\ref{fig:p2d_diag}b shows the resulting downsampling pattern in the spatial-frequency domain. Based on Figure~\ref{fig:p2d_diag}, the benefit of diagonal pilot patterns becomes apparent, as the number of subframes needed to acquire the downsampled CSI matrix, $\bar{\mathbf H}$ at the UE decreases with increasing $D$. For example, the given diagonal size $D=4$ requires 4 subframes (ms) to acquire $\bar{\mathbf H}$, while $D=1$ (i.e., no diagonal pattern or vertical columns of pilots) would require 16 subframes (ms) to acquire $\bar{\mathbf H}$.

To utilize the P2D estimator while using diagonal pilot patterns, it is necessary to account for different pilot matrices, $\mathbf{P}_j$ for $j \in [1,\dots,D]$, used with different antennas. These different pilot matrices result in $D$ different P2D estimators, $\mathbf{Q}^\dag_{j,N_t}$. Algorithm~\ref{alg:p2d-diag} outlines the process for acquiring $\tilde{\mathbf H}$ by applying the P2D estimators to $\bar{\mathbf H}$.

\begin{figure*}[!hbtp]
    \centering
    {
      \fontsize{8pt}{8pt}
      \def\svgwidth{1.0\linewidth}
      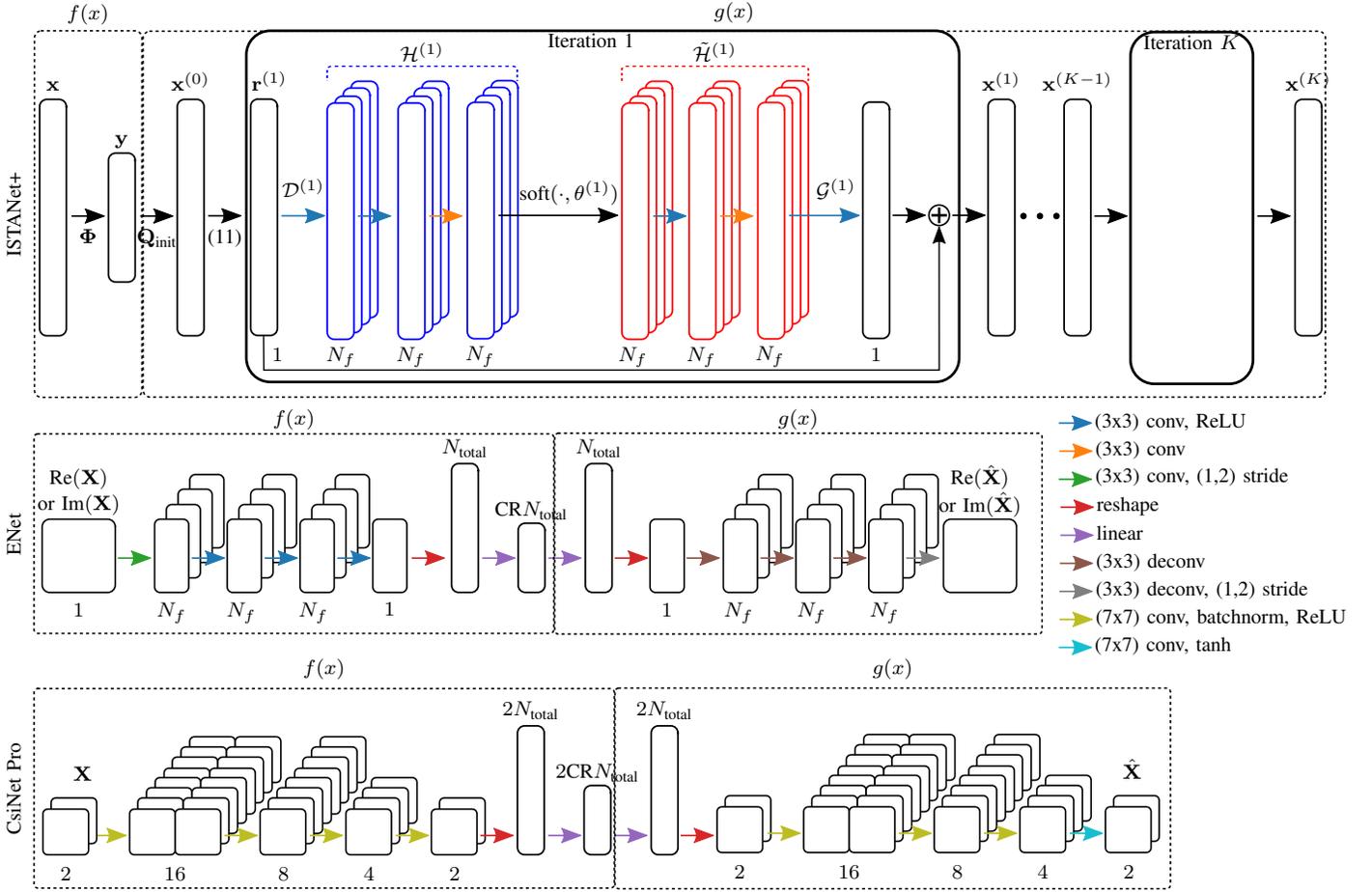
    }
    \caption{Compressive CSI estimation architectures used in this work. $f(x)$ denotes the encoder, and $g(x)$ denotes the decoder. $N_{\text{total}}=N_bN_t$ is the size of the real or imaginary channel.}
    \label{fig:arch_compare}
\end{figure*}

\subsection{Regularization of P2D Estimator}

The pseudoinverse matrices $\mathbf{Q}^\dag_{N_t}$ are typically ill-conditioned given the sparsity of the pilot selection matrices $\mathbf P$. Consequently, the P2D estimator benefits from regularization of the matrix $\mathbf{Q}_{N_t}^H\mathbf{Q}_{N_t}$. This can be done via off-diagonal regularization (ODIR), where all off-diagonal elements are scaled down by a fixed constant. Denote $\mathbf{A}$ as a matrix to be regularized where $A(i,j)$ is the element in the $i$-th row and $j$-th column. The ODIR version of this matrix is
\begin{align}
	A(i,j) &= 
	\begin{cases}
    A(i,j) & i=j \\ 
    \frac{A(i,j)}{1+\delta} & i\neq j \\
    \end{cases}	\label{eq:odir}
\end{align}

\section{Differential Encoding via Learned Compressed Sensing}

To further improve the accuracy of CSI estimation under the P2D estimator, we can exploit the temporal coherence of the channel. Under typical circumstances, the channel does not change substantially for a given window of time, i.e. the coherence interval. Exploiting this coherence is beneficial from an information theoretic point of view \cite{ref:Liu2020MarkovNet}. Denote two subsequent timeslots within a coherence interval as $t_1$ and $t_2$, the entropy of the CSI at $t_1$ as $H(\tilde{\mathbf H}_1)$, and the conditional entropy of the CSI at $t_2$ given $t_1$ as $H(\tilde{\mathbf H}_2|\tilde{\mathbf H}_1)$. Prior work in time-varying CSI estimation has demonstrated that the conditional entropy is always lower than the entropy \cite{ref:Liu2020MarkovNet}, i.e.,
\begin{align}
    H(\tilde{\mathbf H}_2|\tilde{\mathbf H}_1) \leq H(\tilde{\mathbf H}_1).
\end{align}
A reduction in entropy means a reduction in the rate of the compressed feedback, highlighting the utility of differential feedback. Instead of directly encoding/decoding the CSI (e.g., $\hat{\tilde{\mathbf{H}}}_t=g(f(\bar{\tilde{\mathbf{H}}}_t))$), we propose to encode/decode the difference,
\begin{align}
    \bar{\mathbf{E}}_t &= \bar{\tilde{\mathbf{H}}}_t - \check{\tilde{\mathbf{H}}}_t \nonumber \\
                       &= \bar{\tilde{\mathbf{H}}}_t - \gamma\hat{\tilde{\mathbf{H}}}_{t-1}, \label{eq:error-ls}
\end{align}
where $\check{\tilde{\mathbf{H}}}_t = \gamma\hat{\tilde{\mathbf{H}}}_{t-1}$ is the least-squares estimate for $\tilde{\mathbf{H}}_t$ based on the estimate in the previous timeslot, $\hat{\tilde{\mathbf H}}_{t-1}$. We apply the encoding/decoding process to the error term, $\hat{\mathbf{E}}_t=g_t(f_t(\bar{\mathbf{E}}_t))$, and the resulting CSI estimate can be written as 
\begin{align*}
    \hat{\tilde{\mathbf{H}}}_t &= \hat{\mathbf{E}}_t + \gamma\hat{\tilde{\mathbf{H}}}_{t-1}.
\end{align*}
While the feedback is based on the error under the P2D estimator, the network at each timeslot is optimized using the mean-squared error loss function with respect to the error under the ground truth, $\mathbf E_t = \tilde{\mathbf H}_t - \gamma\hat{\tilde{\mathbf H}}_{t-1}$,
\begin{align}
    L_{\text{MSE}} &= \frac {1}{N_{\text{batch}}} \sum_{i=1}^{N_{\text{batch}}} \|\mathbf E^{(i)}_t - \hat{\mathbf E}^{(i)}_t\|_2^2
\end{align}
where $i$ indexes over the $N_{\text{batch}}$ samples of a training batch.

Figure~\ref{fig:markov-p2d} demonstrates the principle of differential encoding used with P2D estimates. Notably, both the BS and the UE need access to a copy of the decoder, $g_t(x)$, in order to derive the error term $\mathbf{E}_t$ based on (\ref{eq:error-ls}). Since both the encoder and the decoder are required on the UE side, we seek to design a differential encoding scheme with a small number of parameters.

\begin{figure}[!hbtp]
    \centering
    \includegraphics{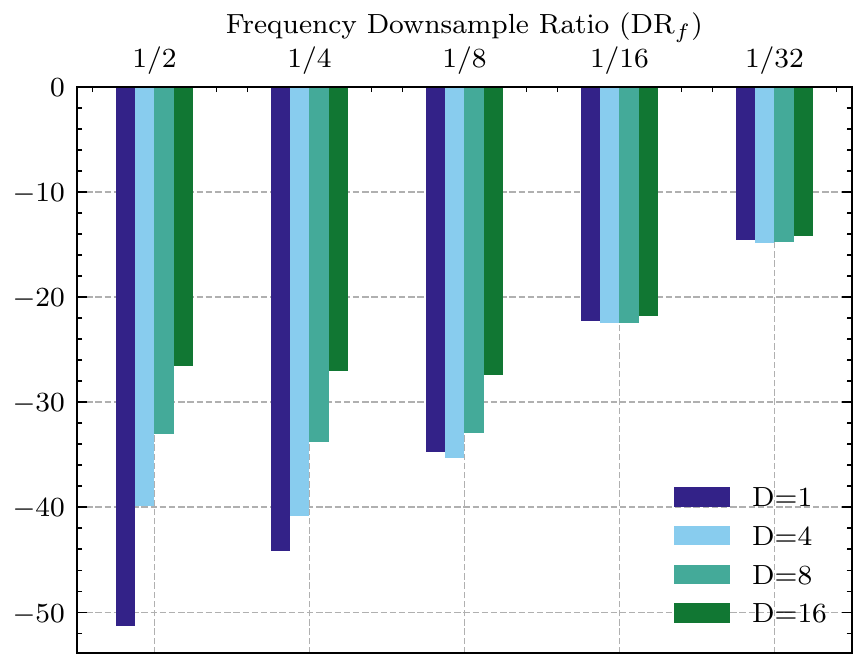}
    \caption{P2D estimation performance under different frequency downsampling ratios ($\text{DR}_f$) and diagonal dimensions ($D$) for the Outdoor COST2100 dataset. Downsampling is done along the frequency axis.}
    \label{fig:outdoor_p2d_init}
\end{figure}

\subsection{CNN Autoencoders for CSI Feedback}

Prior work utilized CNN autoencoders to implement a trainable differential encoding network for CSI estimation \cite{ref:Liu2020MarkovNet}. Using autoencoders in a differential encoding network, each timeslot $t_i$ utilizes a CNN-based encoder ($f_i(x)$) and decoder ($g_i(x)$). Early work in deep learning-based CSI compression concluded that convolutional autoencoders consistently outperformed traditional compressed sensing (CS) approaches \cite{ref:csinet}.

In this work, we investigate two autoencoder networks to realize our differential encoding network. First, we utilize CsiNet Pro \cite{ref:Liu2020SphNet}, an improved version of CsiNet which utilizes a symmetric encoder/decoder structure without residual connections, and ENet \cite{ref:Sun2021ENet}, another symmetric architecture applied independently to the real and imaginary channels to produce a complex-valued matrix. These two networks can be viewed at the bottom of Figure~\ref{fig:arch_compare}.

\subsection{Iterative Optimization Networks for Compressed Sensing-based CSI Feedback} \label{sec:iter-cs}

 
While CNN autoencoders have been dominant in CSI estimation, recent work from image processing has shown promise in using trainable CS algorithms based on CNNs. These works treat iterative CS algorithms as sequential networks by ``unrolling'' them into discrete blocks \cite{ref:yang2016deep, ref:zhang2018ista}. Investigating unrolled CS algorithms for CSI estimation warrants consideration, as CS algorithms can have guaranteed convergence under mild sparsity conditions (in contrast with CNNs autoencoder approaches, which do not have such guarantees). Since CSI data exhibits sparsity in the delay domain, specifying an appropriate compressed sensing approach could provide appreciable performance gains in our differential CSI encoding architecture. 

To exploit the temporal coherence of the MIMO channel, we propose to construct a differential encoding network using an unrolled optimization network based on a trainable version of the iterative shrinkage-thresholding algorithm (ISTA), called ISTANet+ \cite{ref:zhang2018ista}. See the top of Figure~\ref{fig:arch_compare} for a diagram of ISTANet+. Denote the measurement matrix for the ISTANet+ as 
\begin{figure}[!hbtp]
    \centering
    \includegraphics{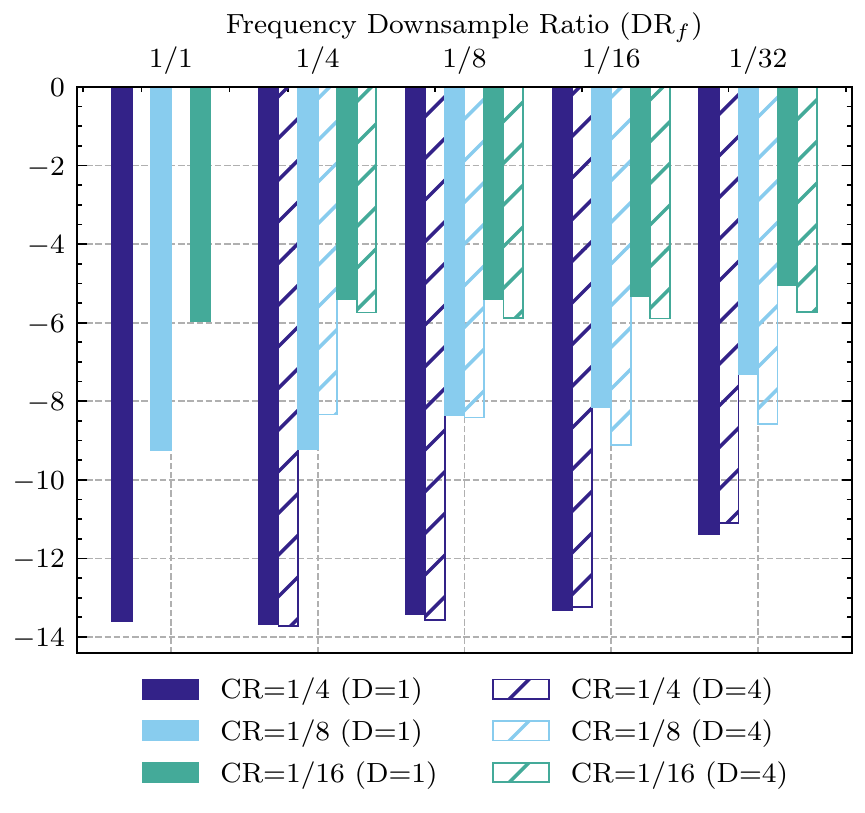}
    \caption{Performance of ISTANet+ for multiple compression ratios using P2D estimates with different downsampling ratios ($\text{DR}_f$) for the Outdoor COST2100 dataset. Non-diagonal pattern ($D=1$) is compared with a diagonal pattern of size $D=4$. Performance for $\text{DR}_f=1/1$, $D=4$ is omitted since it is equivalent to the $\text{DR}_f=1, D=1$ case.} 
    \label{fig:outdoor_drcr_sweep}
\end{figure}
\begin{align}
    \mathbf \Phi \in \mathbf{R}^{N_{\text{total}}\text{CR} \times N_{\text{total}}}.
\end{align}
For compressed sensing approaches, the measurement matrix is the equivalent of the `encoder' for autoencoder approaches, i.e., $f(x)=\mathbf\Phi x$. The `decoder' consists of $K$ iterations of the following update steps,
\begin{align}
    \mathbf{r}^{(k)} &= \mathbf{x}^{(k-1)}-\rho^{(k)}\mathbf{\Phi}^\top(\mathbf{\Phi}\mathbf x^{(k-1)}-\mathbf y) \\
    \mathbf x^{(k)} &= \mathbf{r}^{(k)} + \mathcal{G}^{(k)}\left(\tilde{\mathcal{H}}^{(k)}\left(\text{soft}\left(\mathcal{H}^{(k)}(\mathcal{D}^{(k)}(\mathbf{r}^{(k)}), \theta^{(k)}\right)\right)\right)
\end{align}
where $\mathbf y=\mathbf{\Phi} \mathbf x$, $\mathbf x^{(0)}=\mathbf{Q}^{\text{init}}\mathbf{y}$, and $\mathbf Q_{\text{init}}=\mathbf {XY}(\mathbf{YY}^\top)^{-1}$. The initialization matrix $\mathbf Q_{\text{init}}$ for the training data matrix $\mathbf X = \left[\mathbf{x}_1, \mathbf{x}_2,\dots, \mathbf{x}_{N_{\text{train}}}\right]$ and the training measurement matrix $\mathbf Y = \left[\mathbf{y}_1, \mathbf{y}_2,\dots, \mathbf{y}_{N_{\text{train}}}\right]$. `soft($\cdot$)' denotes the soft threshold function,
\begin{align}
    \text{soft}(x, \theta) &= \text{sign}(x)\text{ReLU}(|x|-\theta).
\end{align}
$\mathcal G^{(k)}, \mathcal D^{(k)},  \mathcal H^{(k)}, \tilde{\mathcal H}^{(k)}$ indicate trainable nonlinear mappings (in this case, CNNs), and $\mathcal H^{(k)}, \tilde{\mathcal H}^{(k)}$ are subject to the symmetry constraint $\mathcal H^{(k)}\circ \tilde{\mathcal H}^{(k)}=\mathbf I$.
\begin{figure}[!hbtp]
    \centering
    \includegraphics{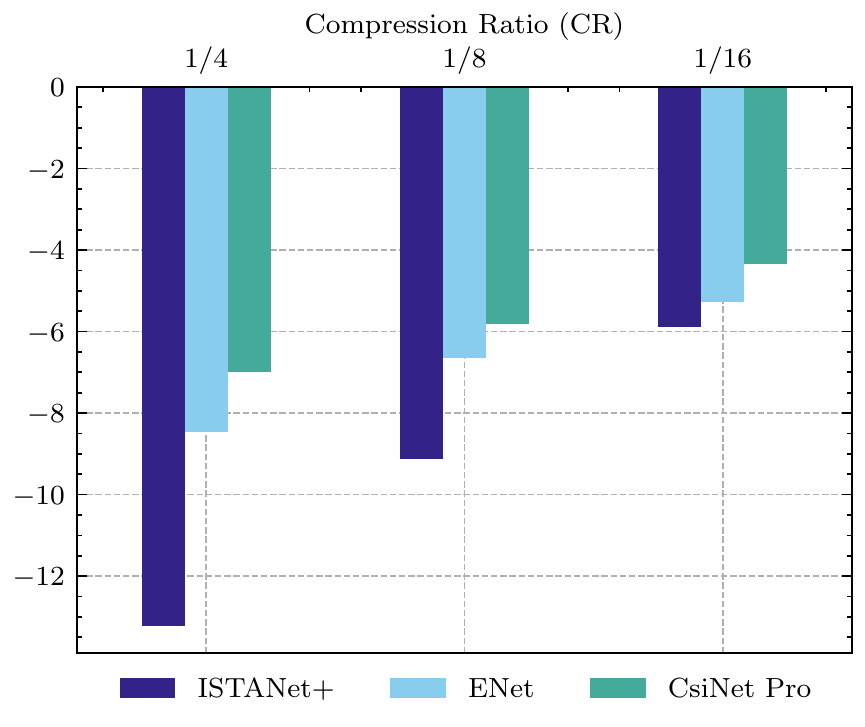}
    \caption{Performance comparison for different feedback compression networks using P2D estimates ($\text{DF}_f=1/16, D=4$) for Outdoor COST2100 dataset.}
    \label{fig:outdoor_net_ablation}
\end{figure}
In the proposed differential encoding scheme, we use an instance of ISTANet+ in the first timeslot, $t_1$, with a large compression ratio such that $\text{CR}_{t_1} \geq \text{CR}_{t_i}$ for all $i > 1$. This choice in compression ratio allows us to initialize the network with a high-quality estimate at the first timeslot. Notably, the training data matrix, $\mathbf X$, differs between timeslots. For the first timeslot, the data vectors $\mathbf{x}_i$ are vectorized versions of the CSI matrices,
\begin{align}
    \mathbf{x}_j &= \text{vec}\left(\bar{\tilde{\mathbf{H}}}^{(j)}_1\right) \text{ for } j\in[N_{\text{train}}]. 
    \label{eq:t1_vec}
\end{align}
However, the data vectors for all other timeslots are vectorized versions of the error matrices,
\begin{align}
    \mathbf{x}_j &= \text{vec}\left(\bar{\mathbf{E}}^{(j)}_{i}\right) \text{ for } j\in[N_{\text{train}}]. 
    \label{eq:ti_vec}
\end{align}
Denote the parameters for ISTANet+ in the $t_i$-th timeslot as $\mathbf{\Theta}_{t_i}=\{\mathcal G^{(k)}, \mathcal D^{(k)},  \mathcal H^{(k)}, \tilde{\mathcal H}^{(k)}\, \theta^{(k)}, \rho^{(k)}\}_{k=1}^{K}$. The loss function is a weighted sum of the MSE and the symmetry constraint, i.e.,
\begin{align}
    L(\mathbf{\Theta}_{t_i}) &= L_{\text{MSE}} + \gamma L_{\text{sym}} \\
    L_{\text{MSE}} &= \frac{1}{N_{\text{batch}}N_{\text{total}}}\sum_{i=1}^{N_{\text{batch}}}\|\mathbf{x}_i^{(K)}-\mathbf{x}_i\|_2^2 \\
    L_{\text{sym}} &= \frac{1}{N_{\text{batch}}N_{\text{total}}}\sum_{i=1}^{N_{\text{batch}}}\sum_{k=1}^{K} \|\tilde{\mathcal{H}}^{(k)}(\mathcal{H}^{(k)}(\mathbf{x}_i)) - \mathbf{x}_i)\|_2^2
\end{align}
where $N_{\text{total}}=N_bN_t$ is the size of the truncated CSI matrix, $K$ is the number of iterations in ISTANet+, and $N_{\text{batch}}$ is the batch size used during training. As denoted in equations (\ref{eq:t1_vec}) and (\ref{eq:ti_vec}), the vectors $\mathbf{x}_i$ depend on the timeslot.

\section{Random Phase Augmentation} \label{sec:random-phase}

\begin{figure}[!hbtp]
    \centering
    \includegraphics{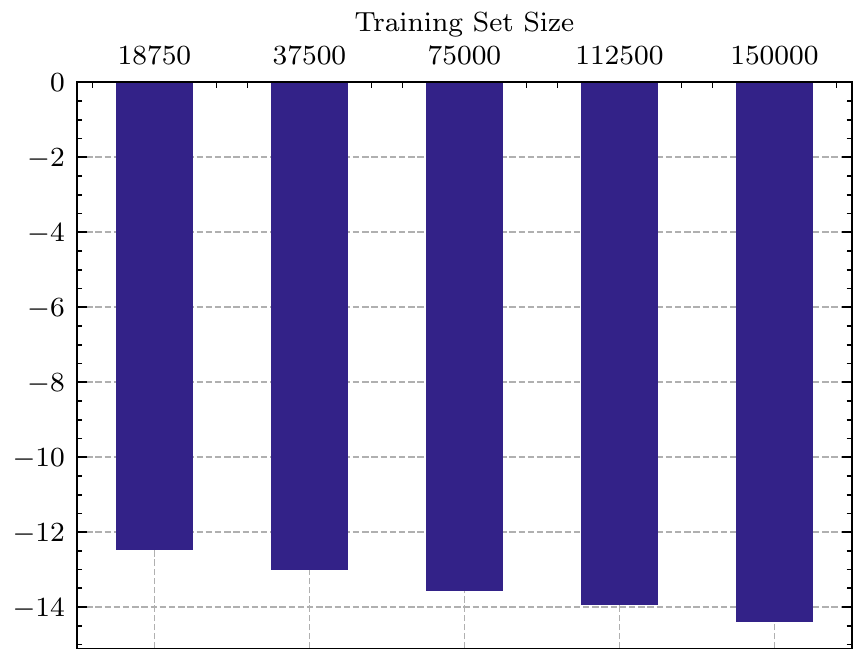}
    \caption{ISTANet+ with P2D performance under different augmented dataset sizes after phase randomization for the Outdoor COST2100 dataset. Downsampling is done along the frequency axis.}
    \label{fig:outdoor_aug_sweep}
\end{figure}

Prior work leveraged the truncated delay domain, which allowed them to save large datasets of truncated CSI matrices. In order to acquire P2D estimates for different values of $\text{DR}_f$ and $D$, we must store the full frequency domain CSI matrices. These full matrices can be prohibitively expensive to store under typical system parameters, meaning we need to use a smaller dataset. Since successful training of deep neural networks depends on a large number of training samples, we utilize a random phase augmentation on our smaller training data. For each sample in the training set, we sample a random phase from a uniform distribution, $\theta \sim \mathcal{U}(-\pi, \pi)$, and we rotate all the elements in a given CSI matrix by this phase,

\begin{align}
    \mathbf{H}^{(i,j)}_{\text{augmented}} &= \mathbf{H}^{(i,j)}e^{-j\theta} \;\forall\; i\in[N_t], j\in[M_f].
\end{align}

 define a \emph{phase augmentation factor}, $N_{\text{phase}}$, which is a multiplicative factor denoting the size of the training dataset after performing phase augmentation. For example, if we begin with a training set of size 5000, then  $N_{\text{phase}}=2$ would result in an augmented dataset of size $10,000$, meaning each sample in the training set is augmented once. More generally, each sample in the training set is augmented $N_{\text{phase}}-1$ times.

\section{Results} \label{sec:results}

\begin{table}[htb]
  \begin{center}
    \caption{Parameters for COST2100 model in this work.}
    \label{tab:cost2100-params} 
    \begin{tabular}{|l|c|}
      \hline 
      \textbf{Environment}      & \textbf{Outdoor} \\ \hline
      Num. gNB Antennas ($N_b$) & 32 \\ \hline
      Truncation Value ($N_t$)  & 32 \\ \hline
      Num. Subcarriers ($N_f$)  & 1024 \\ \hline
      Downsampled Subcarriers ($M_f$)  & $[512,256,128,64]$ \\ \hline
      Carrier Frequency         & 300 MHz \\ \hline
      UE Starting Position      & $400$ m $\times 400$ m \\ \hline
      Num. Channel Samples ($N$)& $10^4$ \\ \hline
    \end{tabular}
  \end{center}
\vspace*{-3mm}
\end{table}

\begin{figure*}[!hbtp]
    \centering
    \includegraphics{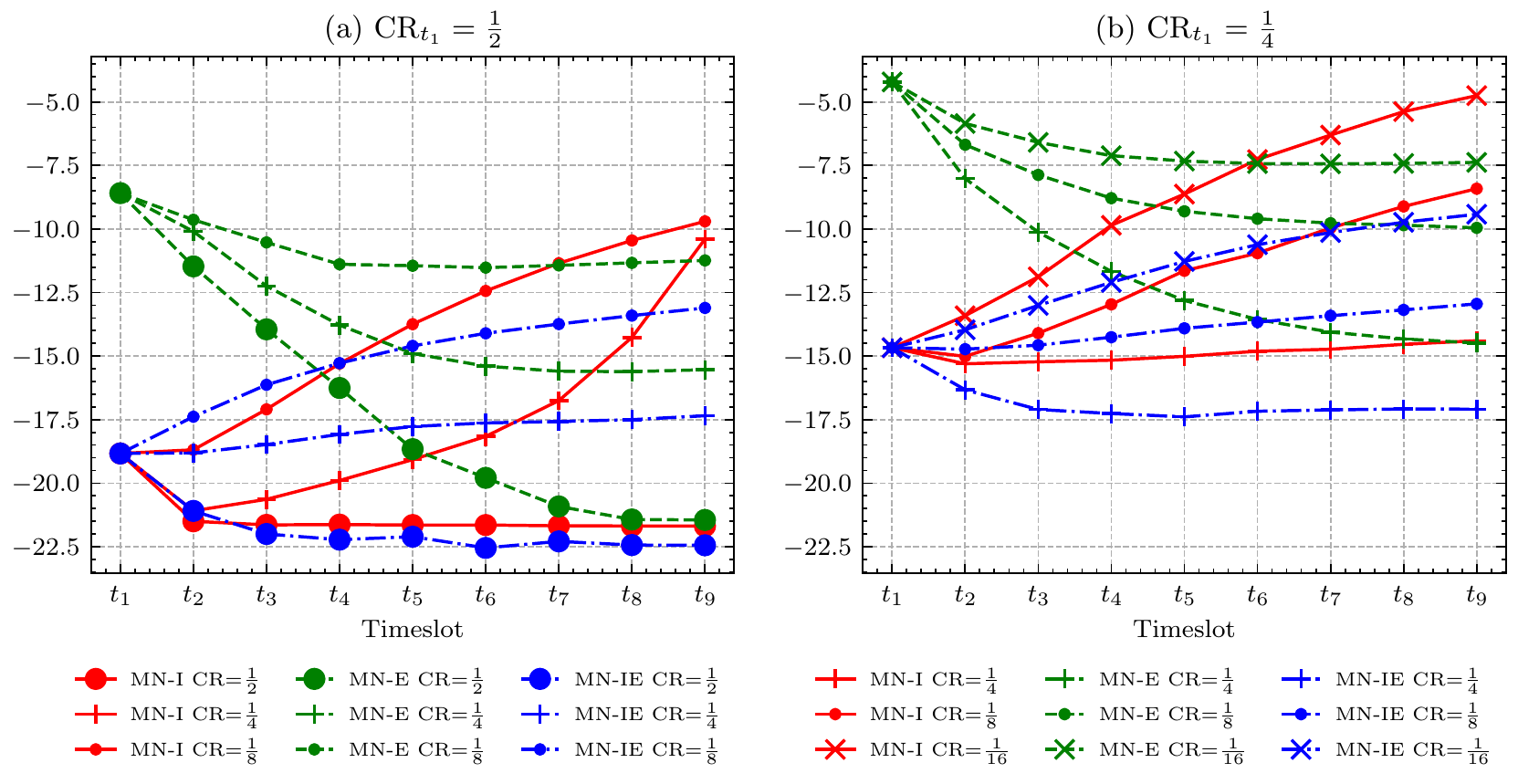}
    \caption{Compressive CSI estimation using differential encoding and the linear P2D estimator ($M_f=128, \text{DR}_f=\frac{1}{8}, D=4$). MarkovNet-ISTA (MN-I), MarkovNet-ENet (MN-E), and MarkovNet-ISTA-ENet (MN-IE) are tested using two different compression ratios in the first timeslot, $\text{CR}_{t_1}\in\left[\frac{1}{2},\frac{1}{4}\right]$.}
    \label{fig:markov-p2d-results}
\end{figure*}

We perform experiments using the COST2100 Model in an Outdoor scenario \cite{ref:cost2100}. Table~\ref{tab:cost2100-params} summarizes the COST model parameters used to generate the Outdoor dataset. Importantly, the number of channel samples in the dataset is lower than the number used in similar works. A smaller dataset is necessary because we store full CSI matrices without truncating any subcarriers, which requires 32 times more space to store. For all networks, we utilize spherical normalization \cite{ref:Liu2020SphNet}, and we test the networks using the following configurations:

\begin{itemize}
    \item \textbf{ISTANet+}: We train the network described in Section~\ref{sec:iter-cs} for 100 epochs using the ADAM optimizer. The network utilizes 32 latent channels, 9 blocks, and a symmetry weight parameter of $\gamma=10^{-3}$. 
    \item \textbf{ENet}: The network hyperparameters are identical to those described in the original paper \cite{ref:Sun2021ENet}. Since the training procedure was not described, we chose one which converged in a reasonable number of epochs (200 epochs, learning rate of $10^{-3}$). As per the original paper, we train the network on the real channel data from the training set, then we report the validation loss by using the network on the real and imaginary channels from the validation set. We utilize $N_f=32$ latent convolutional channels since this configuration achieved the best performance in the original paper.
    \item \textbf{CsiNet Pro}: The hyperparameters and training process are identical to those described in the original paper \cite{ref:Liu2020SphNet}.
\end{itemize}

We use a 75\% (25\%) training (validation) split, yielding 7500 training samples (2500 validation samples). Unless stated otherwise, we augment the training set using $N_{\text{phase}}=10$, yielding an augmented dataset of $75,000$ samples.

\subsection{Accuracy of P2D Estimator}

To provide a bound on the estimation performance at the gNB, Figure~\ref{fig:outdoor_p2d_init} shows the accuracy of the P2D estimator at the UE (i.e., before compression and feedback). The performance of the P2D estimator under multiple diagonal sizes ($D$) is shown. For all tested frequency downsampling ratios ($\text{DR}_f$), the accuracy of the P2D estimator is substantial, with the smallest $\text{DR}_f=\frac{1}{32}$ achieving about -14 dB. For increasing $D$, the error of the P2D estimator increases; however, the difference in performance for different values of $D$ becomes negligible at more aggressive downsampling ratios, $\text{DR}_f \in\left[\frac{1}{16}, \frac{1}{32}\right]$. The accuracy of the P2D estimator implies that it will perform well with compressive CSI feedback networks. 


\subsection{Accuracy of Compressive Networks with P2D Estimates}

In these experiments, we use the P2D estimate as the input to different compressive CSI feedback networks. In this work, we propose to use the unrolled reconstruction network, ISTANet+ \cite{ref:zhang2018ista}, as described in Section~\ref{sec:iter-cs}. In Figure~\ref{fig:outdoor_drcr_sweep}, we assess the performance of ISTANet+ across multiple values of $\text{DR}_f$ and CR. Comparing $\text{DR}_f=\frac{1}{1}$ to $\text{DR}_f=\frac{1}{16}$, the accuracy of ISTANet+ is remarkably stable, increasing negligibly for CR=$\frac 14$ and by only 1 dB for CR=$\frac{1}{16}$.

To provide a baseline for ISTANet+, we also compare the performance of ISTANet+ with two autoencoder-based CSI compression networks, CsiNet Pro \cite{ref:Liu2020SphNet} and ENet \cite{ref:Sun2021ENet}. Figure~\ref{fig:outdoor_net_ablation} shows the performance comparison between all networks for the same $\text{DR}_f$ and $D$. Across all compression ratios, ISTANet+ achieves a better NMSE than the autoencoder approaches.

\subsection{Phase Augmentation Ablation}

Using random phase augmentation as described in Section~\ref{sec:random-phase}, we assess the influence of different sized training sets on validation accuracy. Starting with a training set of size 18750, we augment the dataset by $N_{\text{phase}} \in [2, 4, 6, 8]$, yielding training sets of size 37500, 75000, 112500, and 150000. We train ISTANet+ (CR=$\frac 14$) without P2D (i.e., perfect delay domain data) on each of these training sets, and we report the validation loss on the same 6250 samples. The resulting validation accuracy can be seen in Figure~\ref{fig:outdoor_aug_sweep}. As expected, the accuracy improves appreciably as the size of the augmented training set is increased.

\subsection{Differential Encoding with P2D Estimates}

Figure~\ref{fig:markov-p2d-results} shows the performance of differential encoding when using either ISTANet+ and ENet at each timeslot, which are respectively named MarkovNet-ISTA (MN-I) and MarkovNet-ENet (MN-E). For all versions of MarkovNet, $\text{CR}_{t_1}$ is the compression ratio in the first timeslot and CR is the compression ratio for all following timeslots. ISTANet+ has the benefit of providing accuracy in the first timeslot, while ENet is better at compressing the residual in each following timeslot. Based on this observation, we also test a version of MarkovNet which uses ISTANet+ in the first timeslot then ENet in the following timeslots, which we call MarkovNet-ISTA-ENet (MN-IE). For the networks where $\text{CR}_{t_1}=\text{CR}$, MN-IE can outperform MN-I, indicating that a combination of architectures can be better than a single architecture.

\section{Discussion}

In this work, we present the P2D estimator, a linear estimator for the truncated angular-delay domain CSI based on downsampled spatial-frequency CSI. The P2D estimator provides accurate delay domain CSI based on practical CSI-RS pilot allocations that adhere to the LTE standard. Furthermore, we demonstrate that CSI estimates from the P2D estimator provide a suitable input to trainable CS networks and autoencoder networks. Finally, we propose a differential encoding network, MarkovNet-ISTA-ENet, which combines a trainable CS network with multiple autoencoders to better leverage the high initial accuracy of the former and the error-compressing capabilities of the latter.


\bibliographystyle{IEEEtran}
\bibliography{main.bib}

\begin{thebibliography}{10}
\providecommand{\url}[1]{#1}
\csname url@samestyle\endcsname
\providecommand{\newblock}{\relax}
\providecommand{\bibinfo}[2]{#2}
\providecommand{\BIBentrySTDinterwordspacing}{\spaceskip=0pt\relax}
\providecommand{\BIBentryALTinterwordstretchfactor}{4}
\providecommand{\BIBentryALTinterwordspacing}{\spaceskip=\fontdimen2\font plus
\BIBentryALTinterwordstretchfactor\fontdimen3\font minus
  \fontdimen4\font\relax}
\providecommand{\BIBforeignlanguage}[2]{{%
\expandafter\ifx\csname l@#1\endcsname\relax
\typeout{** WARNING: IEEEtran.bst: No hyphenation pattern has been}%
\typeout{** loaded for the language `#1'. Using the pattern for}%
\typeout{** the default language instead.}%
\else
\language=\csname l@#1\endcsname
\fi
#2}}
\providecommand{\BIBdecl}{\relax}
\BIBdecl

\bibitem{ref:mimo-capacity}
A.~{Goldsmith}, S.~A. {Jafar}, N.~{Jindal}, and S.~{Vishwanath}, ``Capacity
  limits of mimo channels,'' \emph{IEEE Journal on Selected Areas in
  Communications}, vol.~21, no.~5, pp. 684--702, June 2003.

\bibitem{ref:Kaltenberger2010relative}
F.~{Kaltenberger}, H.~{Jiang}, M.~{Guillaud}, and R.~{Knopp}, ``{Relative
  Channel Reciprocity Calibration in MIMO/TDD Systems},'' in \emph{2010 Future
  Network Mobile Summit}, June 2010, pp. 1--10.

\bibitem{ref:mi2017massive}
D.~Mi, M.~Dianati, L.~Zhang, S.~Muhaidat, and R.~Tafazolli, ``Massive mimo
  performance with imperfect channel reciprocity and channel estimation
  error,'' \emph{IEEE Trans. Communications}, vol.~65, no.~9, pp. 3734--3749,
  2017.

\bibitem{ref:Gao2010utilization}
Q.~{Gao}, F.~{Qin}, and S.~{Sun}, ``Utilization of channel reciprocity in
  advanced mimo system,'' in \emph{2010 5th International ICST Conference on
  Communications and Networking in China}, Aug 2010, pp. 1--5.

\bibitem{ref:csinet}
C.~{Wen}, W.~{Shih}, and S.~{Jin}, ``{Deep Learning for Massive MIMO CSI
  Feedback},'' \emph{IEEE Wireless Communications Letters}, vol.~7, no.~5, pp.
  748--751, Oct 2018.

\bibitem{ref:Lu2020CRNet}
Z.~{Lu}, J.~{Wang}, and J.~{Song}, ``Multi-resolution csi feedback with deep
  learning in massive mimo system,'' in \emph{ICC 2020 - 2020 IEEE
  International Conference on Communications (ICC)}, 2020, pp. 1--6.

\bibitem{ref:Hussien2020PRVNet}
M.~Hussien, K.~K. Nguyen, and M.~Cheriet, ``{PRVNet: Variational autoencoders
  for massive MIMO CSI feedback},'' \emph{arXiv}, 2020.

\bibitem{ref:Sun2020AnciNet}
Y.~{Sun}, W.~{Xu}, L.~{Fan}, G.~Y. {Li}, and G.~K. {Karagiannidis}, ``Ancinet:
  An efficient deep learning approach for feedback compression of estimated csi
  in massive mimo systems,'' \emph{IEEE Wireless Communications Letters},
  vol.~9, no.~12, pp. 2192--2196, 2020.

\bibitem{ref:dualnet}
Z.~Liu, L.~Zhang, and Z.~Ding, ``{Exploiting Bi-Directional Channel Reciprocity
  in Deep Learning for Low Rate Massive MIMO CSI Feedback},'' \emph{IEEE
  Wireless Comm. Letters}, vol. 8(3), pp. 889--892, 2019.

\bibitem{ref:csinet-lstm}
T.~{Wang}, C.~{Wen}, S.~{Jin}, and G.~Y. {Li}, ``Deep learning-based csi
  feedback approach for time-varying massive mimo channels,'' \emph{IEEE
  Wireless Communications Letters}, vol.~8, no.~2, pp. 416--419, April 2019.

\bibitem{ref:Liu2020MarkovNet}
Z.~Liu\textdagger, M.~del Rosario\textdagger, and Z.~Ding, ``{A Markovian
  Model-Driven Deep Learning Framework for Massive MIMO CSI Feedback},''
  \emph{IEEE Transactions on Wireless Communications}, pp. 1--1, 2021.

\bibitem{ref:Asplund2020}
\BIBentryALTinterwordspacing
H.~Asplund, D.~Astely, P.~von Butovitsch, T.~Chapman, M.~Frenne,
  F.~Ghasemzadeh, M.~Hagström, B.~Hogan, G.~Jöngren, J.~Karlsson,
  F.~Kronestedt, and E.~Larsson, ``{Chapter 8 - 3GPP Physical Layer Solutions
  for LTE and the Evolution Toward NR},'' in \emph{Advanced Antenna Systems for
  5G Network Deployments}.\hskip 1em plus 0.5em minus 0.4em\relax Academic
  Press, 2020, pp. 301--350. [Online]. Available:
  \url{https://www.sciencedirect.com/science/article/pii/B9780128200469000083}
\BIBentrySTDinterwordspacing

\bibitem{ref:Liu2020SphNet}
Z.~{Liu}, M.~{del Rosario}, X.~{Liang}, L.~{Zhang}, and Z.~{Ding}, ``Spherical
  normalization for learned compressive feedback in massive {MIMO CSI}
  acquisition,'' in \emph{IEEE ICC Workshops}, 2020, pp. 1--6.

\bibitem{ref:Sun2021ENet}
Y.~Sun, W.~Xu, L.~Liang, N.~Wang, G.~Y. Li, and X.~You, ``A lightweight deep
  network for efficient csi feedback in massive mimo systems,'' \emph{IEEE
  Wireless Communications Letters}, 2021.

\bibitem{ref:yang2016deep}
Y.~Yang, J.~Sun, H.~Li, and Z.~Xu, ``Deep admm-net for compressive sensing
  mri,'' in \emph{Proceedings of the 30th international conference on neural
  information processing systems}, 2016, pp. 10--18.

\bibitem{ref:zhang2018ista}
J.~Zhang and B.~Ghanem, ``Ista-net: Interpretable optimization-inspired deep
  network for image compressive sensing,'' in \emph{Proceedings of the IEEE
  conference on computer vision and pattern recognition}, 2018, pp. 1828--1837.

\bibitem{ref:cost2100}
L.~{Liu}, C.~{Oestges}, J.~{Poutanen}, K.~{Haneda}, P.~{Vainikainen},
  F.~{Quitin}, F.~{Tufvesson}, and P.~D. {Doncker}, ``The {COST 2100 MIMO}
  channel model,'' \emph{IEEE Wireless Comm.}, vol.~19, no.~6, pp. 92--99, Dec.
  2012.

\end{thebibliography}

\end{document}